\documentclass[sigconf,natbib=false]{acmart} 

\usepackage{float}
\usepackage{graphicx} 
\usepackage[inline]{enumitem}
\usepackage{calc}
\usepackage{hyperref}

\copyrightyear{2024}
\acmYear{2024}
\setcopyright{rightsretained}
\acmConference[EuroPLoP 2024]{29th European Conference on Pattern Languages of Programs, People, and Practices}{July 3--7, 2024}{Irsee, Germany}
\acmBooktitle{29th European Conference on Pattern Languages of Programs, People, and Practices (EuroPLoP 2024), July 3--7, 2024, Irsee, Germany}\acmDOI{10.1145/3698322.3698444}
\acmISBN{979-8-4007-1683-6/24/07}

\begin{document}
\pagenumbering{gobble}

\clearpage
\title{\centering
AI Future Envisioning with PLACARD 
}
\settopmatter{printfolios=true}

\author{Mary C. Tedeschi}
\orcid{0000-0002-7309-6870}
\affiliation{%
  \institution{Pace University}
  \streetaddress{One Pace Plaza}
  \city{New York}
  \state{NY}
  \country{USA}
  \postcode{10038}}
\email{mtedeschi@pace.edu}  

\author{Paola Ricaurte}
\orcid{0000-0001-9952-6659}
\affiliation{%
  \institution{Tecnologico de Monterrey}
  \streetaddress{Calz México-Xochimilco 4900}
  \city{Tlalpan}
  \state{Ciudad de México}
  \country{Mexico}
  \postcode{14370 CDMX}}
\email{pricaurt@tec.mx}

\author{Sridevi Ayloo}
\orcid{0009-0004-6065-4847}
\affiliation{%
  \institution{New York City College of Technology}
  \streetaddress{300 Jay St}
  \city{Brooklyn}
  \state{NY}
  \country{USA}
  \postcode{11201}}
\email{sayloo@citytech.cuny.edu}

\author{Joseph Corneli}
\orcid{0000-0003-1330-4698}
\affiliation{%
  \institution{Hyperreal Enterprises Ltd.}
  \streetaddress{272 Bath Street}
  \city{Glasgow}
  \state{Scotland}
  \country{UK}
  \postcode{G2 4JR}}
\email{joseph.corneli@hyperreal.enterprises}  
\affiliation{%
  \institution{Oxford Brookes University}
  \streetaddress{Headington Rd}
  \city{Headington}
  \state{Oxford}
  \country{UK}
  \postcode{OX3 0BP}}  
\email{jcorneli@brookes.ac.uk}

\author{Charles Jeffrey Danoff}
\orcid{0000-0002-7086-3587}
\affiliation{%
  \institution{Mr Danoff’s Teaching Laboratory}
 \streetaddress{PO Box 802738}
 \city{Chicago}
 \state{IL}
  \country{USA}
  \postcode{60680}}
\email{contact@mr.danoff.org}
  
\author{Sergio Belich}
\orcid{0009-0002-9367-3310}
\affiliation{%
  \institution{New York City College of Technology}
  \streetaddress{300 Jay St}
  \city{Brooklyn}
  \state{NY}
  \country{USA}
  \postcode{11201}}
\email{sbelich@citytech.cuny.edu}

\renewcommand{\shortauthors}{Tedeschi et al.}

\setcounter{section}{0}

\begin{abstract}
At EuroPLoP 2024 Mary Tedeschi led the “AI Future Envisioning with PLACARD” focus group in Germany. Three conference attendees joined in the room while Sridevi, Paola, and Charles co-facilitated remotely via a web conference. The participants were introduced to a Futures Studies technique with the goal of capturing envisionments of Artificial Intelligence (AI) going forward. To set an atmosphere a technology focused card game was used to make the session more interactive. To close everyone co-created a Project Action Review to recap of the event to capture learnings that has been summarized in this paper. The Focus Group was structured based on lessons learned over six earlier iterations.
\end{abstract}

\ccsdesc[500]{Applied computing~Education}
\ccsdesc[500]{Applied computing~E-learning}

\keywords{AI, distance learning, educational technology, learning management system, online education/teaching, patterns, student engagement}

\received{8 March 2024}
\received[Revised]{14 August 2024}
\received[Accepted]{30 September 2024}

\maketitle

\section{Introduction}

Over the last several years, Artificial Intelligence (AI) has become part of everyday life for many computer and smartphone users.  However, the governance and future directions for these technologies are unclear.  Motivated by this state of affairs, we wanted to convene a discussion that would invite pattern language experts to think creatively together about the future of AI technologies.  We have developed a methodology for structured discussions which builds on design pattern methods, and augments them with techniques from other professional domains.  We deployed this combined methodology in our "AI Future Envisioning" workshop at EuroPLoP 2024. Although the workshop was only small-scale, we have envisioned it as a pilot for subsequent large-scale discussion and innovations (both social and technical) in the way we think about AI technologies.
\\ \\
The session began with basic introductions and definitions, using a PowerPoint deck \cite{1}. On slide 4 (Fig. 1) Christopher Alexander's \cite{2}\cite{3} design patterns were used as a point of entry into the vision of Causal Layered Analysis (\textbf{CLA}). A pioneer of Design Pattern Languages (\textbf{DPL}) was chosen for the pattern conference attendees to find a connection or bridge to the new (for them) Futures Studies technique. The participants then played {\textquotedbl}The Oracle for Transfeminist Technologies{\textquotedbl} card game as a creative thinking tool to source ideas for the future of Artificial Intelligence. Participants further used the \textbf{CLA} framework to refine their concepts on the flip board. At the end, a Project Action Review (\textbf{PAR}) was implemented to document the outcomes. We first introduced the synthesis of these methods (\textbf{CLA + PAR + DPL = PLACARD}) in our 2021 paper Patterns of Patterns \cite{2}. 

\begin{figure}[H]
 \includegraphics[width=2.1in,height=1.2in]{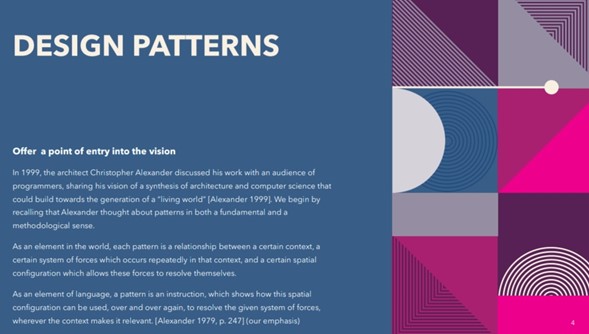} 
    \caption{Slide 4 from presentation used for the Focus Group}
\end{figure}
The session was led by Mary onsite in Kloster Irsee Germany. Paola, Sridevi, and Charles co-facilitated remotely from North America through a Jitsi web conference call via a laptop in the room. We were in the 17:00 – 18:30 CEST time slot on July 5th, 2024. After explaining the objectives (2), process (3), and outcomes (4) this paper includes a background section (5) showing how this session evolved over six workshops, beginning with PLoP 2021, followed by a summary discussion (Section 6) that contextualises this work.  Our conclusion (Section 7) highlights directions we hope to pursue in further work.
\\

\section{Objectives}

Our main objective was to facilitate a CLA exercise to envision the future of AI. Each participant used a value card; an object card; a bodies and territories card; and a situation card. The cards were distributed and participants were asked to read their cards. 
\newline \\
A value card represents a transfeminist value.  The participants were asked what the value meant to them. Object cards represent everyday objects. Bodies/Territories cards encourage us to recognize the importance of situated and embodied knowledge. We asked the participants to reflect on who they are. What body are you in? Where do you live? What privileges and burdens do your body and place provide? The situation card provides you with a situation that you need to deal with. They can be a fun way to focus your creative energy.
\newline \\
Heading into the focus group we hoped our PLACARD envisionments would yield fruitful insights into AI that will aid humanity. We were particularly interested in exploring how AI could be used, adapted, and governed in ways that bring the greatest benefits to the most people. At the same time we were concerned that, currently, AI engineering is not accessible to many people, and that although AI systems can in many cases be widely used, they may be “extractive by design” insofar as they serve the profit motives of the companies that have developed them. 
\newline \\
We hoped to build on our experience in the previous parts of this workshop to discuss what we think about these concerns, and use PLACARD to envision some paths forward. Our goal was to create a futuristic technology first without AI and then revise it with AI.
\newline \\
Two authors attended the "CheriSharing: Sharing Cherished Qualities of Pattern Languages (Focus Group 2)" \cite{4} at PLoP 2023 in Allerton Park that was developed “to ‘share’ the goodness of a certain theme through conversations and ‘cherish’ it together.” Another desired outcome of ours was to have a similar level of wonderful participation and interaction from attendees.

\section{Process}

The session was divided into two 20-minute segments. Mary worked with attendees in the room in Germany while Paola, Sridevi, and Charles supported remotely from North America. The room included:

\begin{itemize}[noitemsep,topsep=8pt]
\item Flip chart
\item Projector, internet connection, laptop, and screen so remote co-facilitators could join 
\item The Oracle for Transfeminist Technologies card deck \cite{5} as prompts for creativity
\end{itemize} 

We gave illustrative examples of CLA via the projected slides and verbally with the “Parking space” example \cite{6} from Dr.
Inayatullah: 
\\

\textit{“However, CLA orders the scenarios in vertical space. For example, taking the issue of parking spaces in urban centers can lead to a range of scenarios. A short term scenario of increasing parking spaces (building below or above) is of a different order than a scenario which examines telecommuting or a scenario which distributes spaces by lottery (instead of by power or wealth) or one which questions the role of the car in modernity (a carless city?) or deconstructs the idea of a parking space, as in many third world settings where there are few spaces designated ‘parking’.” }

\subsection{Example Scenarios}

The “Transfeminist Oracle” cards provided participants manipulative creative structures around values, situations, and objects. Initially, we did not plan to use the situation cards, but one participant asked, "What problem are we trying to solve?" This question guided us to incorporate situation cards, making the exercise more engaging and purposeful.
\newline \\
\begin{figure}[H]
 \includegraphics[width=2in,height=1.31in]{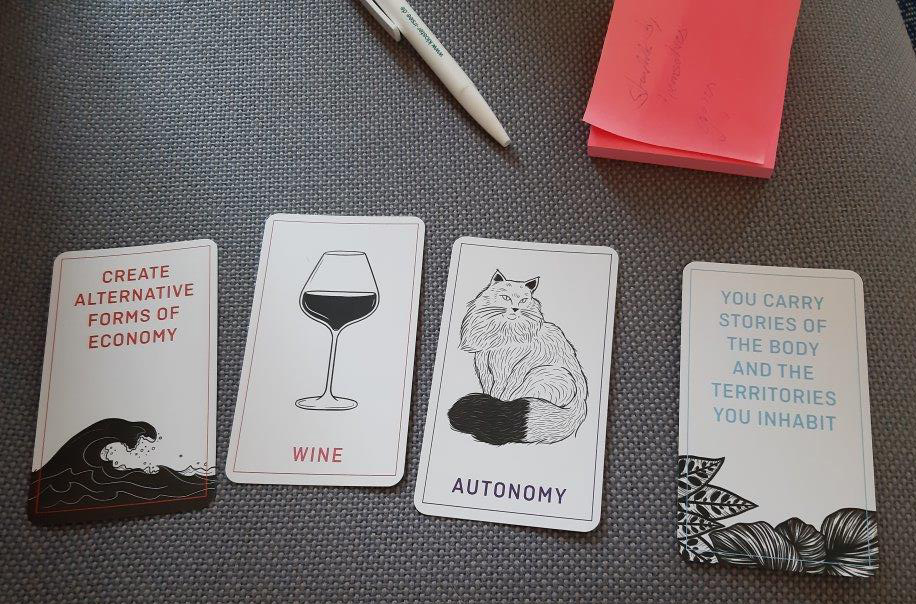} 
    \caption{Credit to Monika}
\end{figure}

One participant had the \textbf{Guitar} card (object) with \textbf{Empathy} (value) aiming to grant unrestricted access to information. Another had the \textbf{Wine} card (object) with \textbf{Autonomy} (value) focused on creating alternative forms of economy. The third participant had the \textbf{Bicycle} card (object) with \textbf{Security} (value) addressing diffuse power and diversified governance.
\newline \\
Each participant shared the same card named \textbf{“Body/Territory”} card, which states “you carry stories of the body and territories you inhabit”. The objects served as metaphors: the bicycle represented an electric car or electric bicycle, the guitar symbolized music, and wine was something delicious.

\subsection{Activities}

Participants considered the CLA, represented via a diagram with concentric rings, and wrote their thoughts on post-it notes, and posted them on a copy of the diagram on a flip chart.
\\ 
\begin{figure}[H]
 \includegraphics[width=1.75in,height=1.813in]{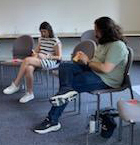} 
    \caption{EuroPlop ‘24 Workshop - participants writing, photo by Mary Tedeschi}
\end{figure}

\vspace{10pt}
\begin{figure}[H]
 \includegraphics[width=2.0in,height=1.86in]{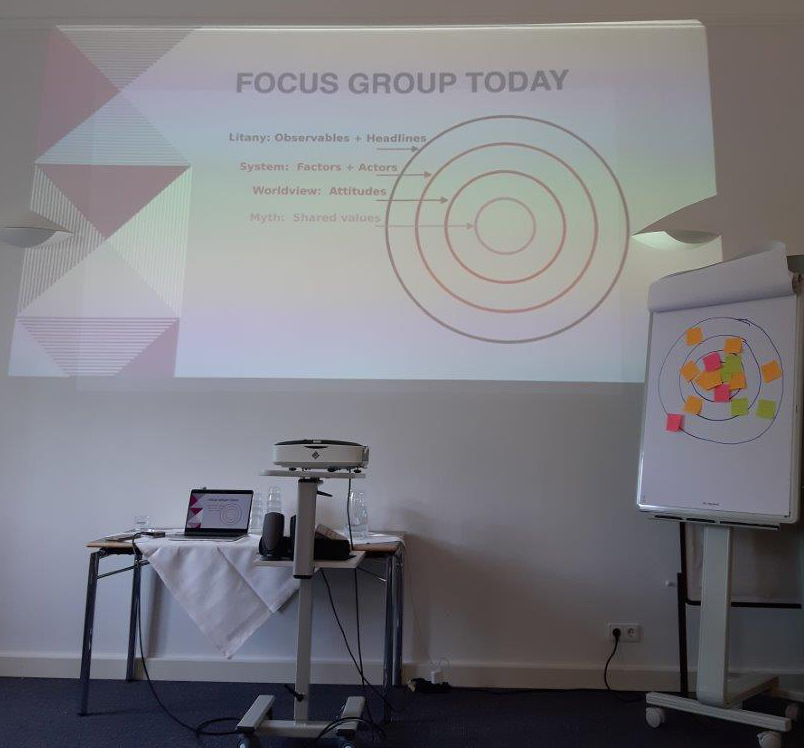} 
    \caption{ EuroPLoP ‘24 - Presentation slide showing CLA and Flip Chart with Post-it Notes from Participants plus online Co-Facilitators, photo by Monika}
\end{figure}

\vspace{10pt}
\begin{figure}[H]
 \includegraphics[width=2in,height=2.25in]{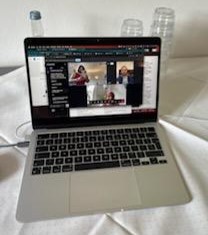} 
    \caption{EuroPlop ‘24 Co-Facilitators on Jitsi, photo by Mary}
\end{figure}

\section{Outcomes}

We successfully completed the planned activities with five minutes to spare. The online co-faciliatators  provided support but did not actively engage in the exercises. 

\subsection{PAR (Project Action Review)}

We closed with a collective PAR where the groups shared back what they discussed and learned. We collaboratively answered the five questions \cite{7}. The initial notes were taken in the Jitsi Chat \cite{8}. You can see one screenshot in the figure below. A consolidated revision hoping to capture their essence is below. The five questions are:
\begin{enumerate*}
    \item \emph{Review the intention: what do we expect to learn or make together?}
    \item \emph{Establish what is happening: what and how are we learning?}
    \item \emph{What are some different perspectives on what’s happening?}
    \item \emph{What did we learn or change?}
    \item \emph{What else should we change going forward?}
\end{enumerate*}
From attendees’ responses to these questions we heard back that they perceived the collective intention in terms of coming to a shared understanding of the PLACARD method and its constituents such as CLA, via examples.  Attendees found the experience enlightening, and suggested that the method could help people think creatively and bring companies out of their “business as usual” way of operating. They found the examples of previous practical use helpful, and felt that the cards came in handy. In addition to gaining experience with the methods, they left with more confidence about asking “what if?” and imagining different futures.
\\ \\

\begin{figure}[H]
 \includegraphics[width=2in,height=2.2in]{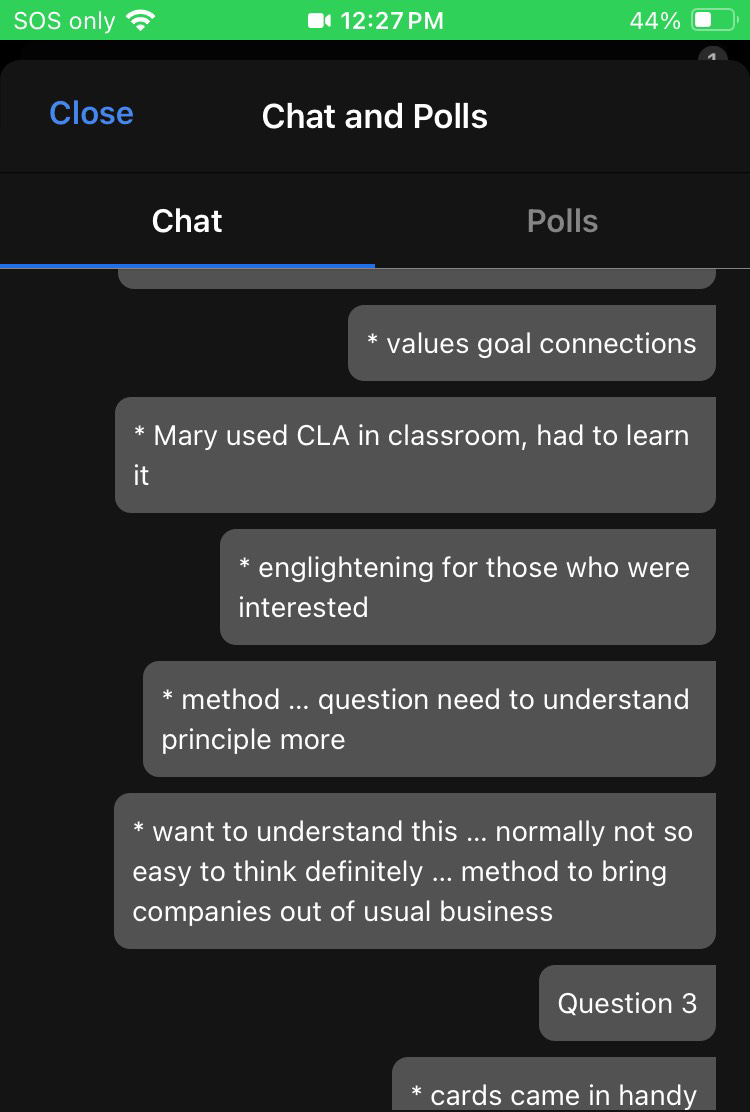} 
    \caption{Screenshot of Jitsi chat with live notes of second and third PAR questions with answers.}
\end{figure}

\subsection{Questionnaire Results}

Following the example of the EuroPLoP 2023 CheriSharing Workshop: Sharing Cherished Qualities \cite{4}, we sent participants a questionnaire about their experiences. It had fourteen questions and the last five were adapted from the PAR.
\\ \\
We received responses from all three participants, for a 100\% response rate. Every respondent had experience attending Hillside PLoP events in the past, so they had familiarity with patterns and focus groups. The anonymous respondent joined the Focus Group “because the title seemed interesting and I also wanted to support the people involved :)” while Tsvetelina Plummer joined because it “Sounded very unfamiliar to me, so it was interesting to learn what you do”. The reason Prof. Dr.-Ing. Monika Blattmeier attended because “I was interested in how the future of AI can be visualized.”
\\ \\
Tsvetlina wrote that she learned “An interesting creativity-sparking exercise” and the other respondent wrote  “I learned a new way to organize ideas and knowledge with PLACARD and its components”. Prof. Dr.-Ing. Monika Blattmeier learned “about Causal Layered Analysis …  which I find very inspiring. Secondly, I was excited about the cards we worked with and came up with new ideas. These cards allowed a complex thinking concept, as PLACARD seems to be, become applicable for practitioners.” Additionally, Tsvetilina plans to use what they learned going forward “I’ll add the new techniques I learned into my toolbelt of processes about learning” as does Prof. Dr.-Ing. Monika Blattmeier “I would also like to use this method for my own project, the development of a knowledge management system.”
\\ \\
All respondents had suggestions for us to consider with future versions of the workshop along the same lines ”I felt that the delivery of the focus group could have been more cohesive and structured by having an overview at the beginning, at points it felt like a new idea was shown to us a bit out of the blue.” Similarly, Prof. Dr.-Ing. Monika Blattmeier wrote “It was hardly possible for me to understand your, in my view, promising PLACARD concept in the focus group. I was not prepared for it, did not know your theory and it was difficult for me to follow the context of the PLACARD concept presented by you in the relatively short time frame of the focus group.” Finally, Tsvetlina suggested “Compress the definitions and textual information into a handout for participants to have after the workshop (for their reference later). Instead put graphics as visual aid and only little texts on the slides – It was very hard to try to follow both the audio and visual tracks. You could also try a version where you start with the exercise and then present a bit of theory – I think that could work as well as participants will have the activity as [an] anchor in their brains when you present the theory later”. We are very thankful for them to take the time to complete our questionnaire and intend on using their feedback to improve future workshops.
\\ \\


\section{Background}

One of the authors of this paper attended another AI focus group “ChatGPT Prompt Engineering for Pattern Writers: Unlock Your Potential” \cite{9} by Stefan Holtel at EuroPLoP 2023. We were curious about how a PLACARD workshop on the future of AI would intersect to extend some of these previous conversations. 
\\ \\
We have previously used PLACARD in institutions and at conference workshops, including AsianPLoP 2024 (Focus Group 4) \cite{10}. Charles led that event onsite in Fujisawa while Sridevi and Mary co-facilitated from New York via Jitsi. We introduced CLA visually, via the iceberg metaphor with this iceberg diagram \cite{11}. Rather than AI, the technique was used to collectively reflect on the conference itself. During the PAR discussion we noted the benefits of paper workshops plus how sometimes people have so much energy for the event itself, but then burn out afterwards so it is difficult to follow up with new connections on new projects. Going through the CLA levels the reflections included that the pattern community in Japan and Asia is smaller than other regions and that AsianPLoP is much bigger than just far East Asia. At the myth level we discussed that pattern languages themselves are a language that transcends human languages: if a pattern itself is good and useful, it does not matter whether it is written in Japanese or English.

\vspace{30pt}

\subsection{2021 - 2024 Versions of Focus Group}

An earlier antecedent of these workshops was a reading of design fiction we did with participants at the Anticipation 2019 \cite{12}. We got feedback from a participant that they did not like to have scripted lines. Since then we have focused on interaction during our PLACARD workshops. Our methods have continued to be remixed over the years (see Table 1). The table was adapted from one used in Sawami’s AsianPLoP paper \cite{13}. Common to all these settings are the need for a rapid training in the methods, so that people can begin to use them after very brief instruction.
\\ \\
This table only lists the public-facing events; other applications of some or all of the methods include sessions with students in Mary Tedeschi’s classroom (carried out over 3 successive iterations of her course), an Oxford Brookes away day about open research, and a workshop with officers of the Oxfordshire County Council.

\vspace{30pt}

\subsection{Other Uses of the PLACARD Method}

The Emacs Research Group [2] conducted a PAR at the end of each of their weekly meetings.  After every ten meetings or so, a CLA was developed (or revised) using bullet-points from the PAR as evidence.  The Peeragogy project experimented with using a DPL as a project management system, by compartmentalizing their collective todo-list as "Next Steps" attached to each of their collected patterns [27].

\onecolumn 

\begin{table*}
  \caption{PLACARD Workshops}
  \centering
  \renewcommand{\arraystretch}{1.5}  
  \begin{tabular}{|p{0.3in}|p{1.4in}|p{1.4in}|p{1.4in}|p{1.4in}|}
    \hline
    \textbf{Year} & \textbf{Event} & \textbf{Modality} & \textbf{Tools} & \textbf{PAR} \\
    \hline
    \vspace{1pt}2021\vspace{1pt} & \vspace{1pt}Oxford Brookes Creative Festival \cite{14}\vspace{1pt} & \vspace{1pt}Online (Zoom)\vspace{1pt} & \vspace{1pt}"Flaws of the Smart City" cards \cite{15}\vspace{1pt} & \vspace{1pt}No\vspace{1pt} \\
    \hline
    \vspace{1pt}2021\vspace{1pt} & \vspace{1pt}PLoP: "Workshop: Flaws of the Cool City" \cite{16}\vspace{1pt} & \vspace{1pt}Online (Discord) \cite{17}\vspace{1pt} & \vspace{1pt}Flaws of the Cool City game\vspace{1pt} & \vspace{1pt}Yes; ``\href{https://youtu.be/i4SMTlpP2uw?si=IaYWuD-4pH3adpbK}{Overall evaluation of our workshop}''\vspace{1pt} \\
    \hline
    \vspace{1pt}2021\vspace{1pt} & \vspace{1pt}Connected Learning Symposium \cite{18}\vspace{1pt} & \vspace{1pt}Online\vspace{1pt} & \vspace{1pt}No\vspace{1pt} & \vspace{1pt}No\vspace{1pt} \\
    \hline
    \vspace{1pt}2022\vspace{1pt} & \vspace{1pt}Anticipation\vspace{1pt} & \vspace{1pt}In Person\vspace{1pt} & \vspace{1pt}Posters\vspace{1pt} & \vspace{1pt}Yes; See our Case Study in Patterns of Patterns II \cite{19}\vspace{1pt} \\
    \hline
    \vspace{1pt}2023\vspace{1pt} & \vspace{1pt}Wikimania Singapore \cite{20}\vspace{1pt} & \vspace{1pt}Hybrid\vspace{1pt} & \vspace{1pt}Our own cards\vspace{1pt} & \vspace{1pt}YouTube video of session \cite{21}\vspace{1pt} \\
    \hline
    \vspace{1pt}2024\vspace{1pt} & \vspace{1pt}AsianPLoP\vspace{1pt} & \vspace{1pt}Hybrid In Person with Online (Jitsi) Co-Facilitators\vspace{1pt} & \vspace{1pt}PLACARD\vspace{1pt} & \vspace{1pt}Yes\vspace{1pt} \\
    \hline
    \vspace{1pt}2024\vspace{1pt} & \vspace{1pt}EuroPloP\vspace{1pt} & \vspace{1pt}Hybrid In Person with Online (Jitsi) Co-Facilitators\vspace{1pt} & \vspace{1pt}PLACARD \& Transfeminist Oracle Cards\vspace{1pt} & \vspace{1pt}Yes\vspace{1pt} \\
    \hline
  \end{tabular}
\end{table*}

\vspace{4\baselineskip}
\begin{figure*}[h]
 \centering
 \includegraphics[width=400px]{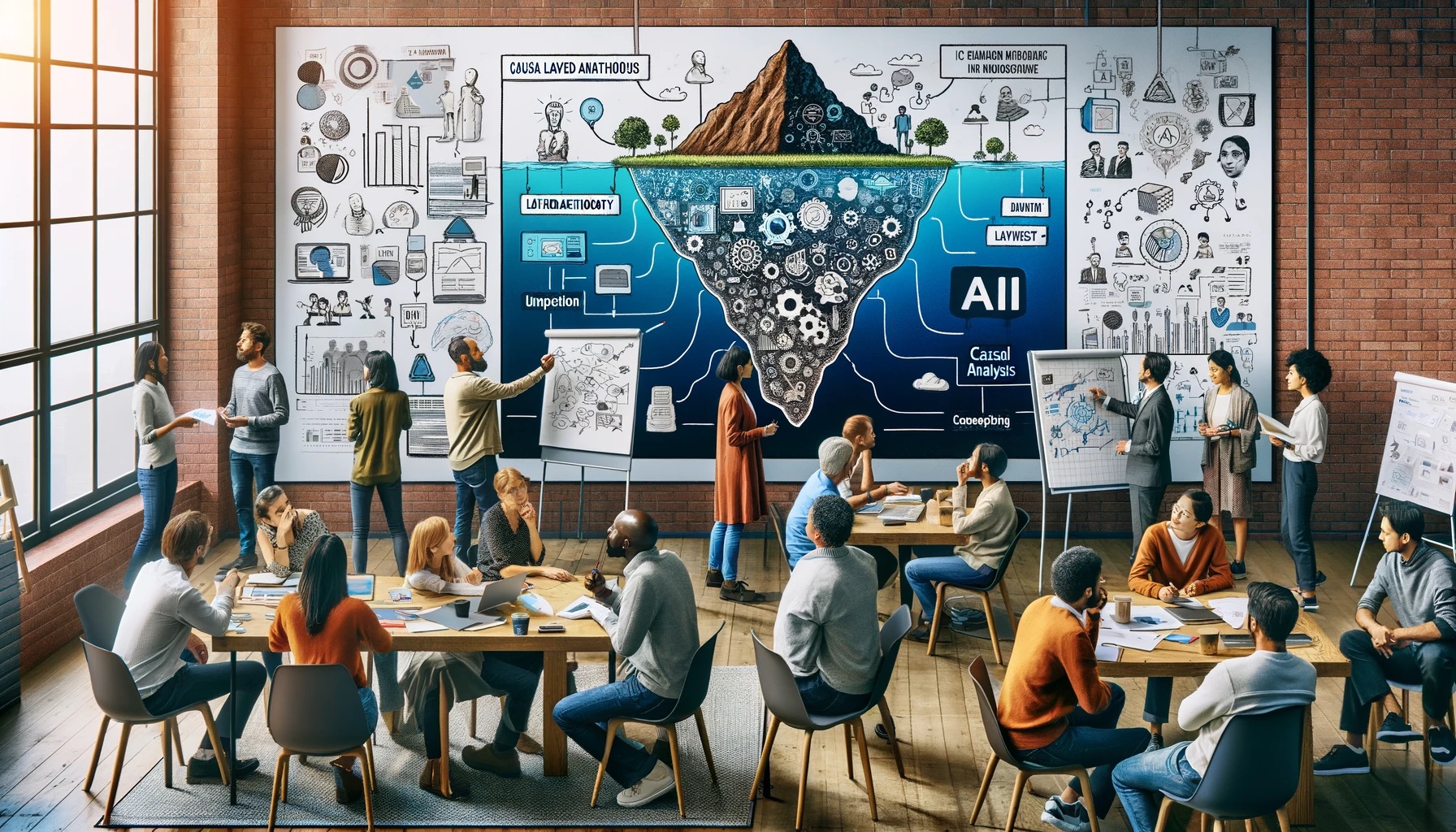} 
   \caption{Image by Open AI DALL$\cdot $E 3 after reading about this focus group.}   
\end{figure*}

\nobreak
\twocolumn

\section{Discussion}

CLA, DPL, and Action Reviews have all been widely used separately in other settings.  However, there is a difference between sharing methods in a way that can be readily replicated by participants on their own, and simply using the methods to gather data. Similarly when participants already have a shared background (as in PLoP settings) more can be done quickly when new methods are introduced.
\\

\subsection{Connections to Peeragogy} 

In our work on a theory of peer learning (peeragogy/paragogy) we elaborated a number of patterns \cite{22} and principles \cite{23} which describe how people create well-functioning peer learning environments. Systems like PLACARD distill this knowledge into operational terms, and can be used primarily for social innovation. Within this workshop, we hoped to engage participants in innovative thinking innovations around current — and future — technical innovations.  While our conclusions about AI are still forming, one take-away from this process is that we need to find more ways to bring media interactions into conversations with each other.  
\\ \\
For example, how could we create “dialogue” across the above workshop interactions? What infrastructure would be needed to be able to make different sets of patterns and proto-patterns (e.g., design prompts) interoperate? Could one approach be making a method for interoperable card games such as the Transfeminist Oracle cards, “Analogia” from MIT \cite{24}, and “Collaboration Cards” \cite{25} from Iba Lab?  Will AI tools be able to help with these challenges to build a new “Discovery Machine” akin to the Hubble Space Telescope \cite{26}? We worked with DALL-E to envision possible future vision of AI with PLACARD, as seen in Fig. 7. 
\\ \\
It is in principle possible to train an AI system based on a collection of existing pairs of patterns to create custom AI based on the pattern language literature. A naive approach building on the existing literature would not respect creators’ copyright; however we do want to respect creators’ copyright. Perhaps we could build a system together with other workshops and a copyright waiver or license, whereby future workshop participants could agree to share their ideas in a way that would allow them to be fed into future models.



\section{Conclusion}
The focus group effectively facilitated creative and critical thinking, leveraging metaphors and participatory methods to explore futuristic technologies and social values. The outcome ended up being more about the methods than the application of envisioning the future of AI. We are pleased that the participants shared in the questionnaire that they learned something new from the Focus Group. We plan to use their suggestions to help co-create our next Focus Groups. We are excited about the way PLACARD was used to have new conversations and are interested in ways it can bring different learnings together. As for envisioning the future of AI, we are considering possibilities, including a new pattern language and/or revising our mini handbook  \cite{27}.

\end{document}